\documentstyle[aaspp4,epsf]{article}

\begin{document}

\title{Analyzing data from DASI}
\author{Martin White${}^1$, John Carlstrom${}^2$,
Mark Dragovan${}^3$, S.W.L. Holzapfel${}^4$}
\affil{
${}^1$Harvard-Smithsonian Center for Astrophysics\\
60 Garden St, Cambridge, MA 02138\\
${}^2$Department of Astronomy and Astrophysics, University of Chicago\\
5621 South Ellis Ave, Chicago IL 60637\\
${}^3$Jet Propulsion Laboratory, Pasadena, CA91109\\
${}^4$University of California, Berkeley, CA 94720}
\authoremail{mwhite@cfa.harvard.edu}

\begin{abstract}
\noindent
\rightskip=0pt
We present an optimized layout of horn positions for the Degree Angular
Scale Interferometer (DASI), which provides good coverage of the $u$-$v$
plane and 3-fold symmetry.
We investigate how an optimal subspace filtering analysis of a single
pointing of DASI could be used to determine the anisotropies in the cosmic
microwave background over a range of angular scales near $30'$.
We discuss the complementarity between the angular scales probed by DASI
and NASA's {\sl MAP\/} satellite at 30GHz and strategies for imaging the
sky at these frequencies.
Finally, we quantitatively assess how well the angular power spectrum can
be recovered by deconvolution of linear combinations of squares of the
DASI visibilities.
\end{abstract}


\rightskip=0pt
\section{Introduction}

The study of the Cosmic Microwave Background (CMB) anisotropy holds the
promise of answering many of our fundamental questions about the universe
and the origin of the large-scale structure
(see e.g.~Bennett, Turner \& White~\cite{BenTurWhi}).
The advent of low-noise, broadband, millimeter-wave amplifiers
(Popieszalski~\cite{Pop}) has made interferometry a particularly attractive
technique for detecting and imaging low contrast emission, such as anisotropy
in the CMB.
An interferometer directly measures the Fourier transform of the intensity
distribution on the sky.
By inverting the interferometer output, images of the sky are obtained which
include angular scales determined by the size and spacing of the individual
array elements.

In an earlier paper (White et al.~1998, hereafter \cite{BigPaper}) we outlined
a formalism for interpreting CMB anisotropies as measured by interferometers.
In this paper we extend this analysis to consider an efficient method of
analyzing the data that would be obtained from a series of uncorrelated
pointings of an interferometer.
In particular we examine what the upcoming Degree Angular Scale
Interferometer\footnote{More information on DASI can be found at
{\tt http://astro.uchicago.edu/dasi}.} (DASI, Halverson et al.~\cite{HCDHK})
experiment may teach us about cosmology. 

DASI is an interferometer designed to measure anisotropies in the CMB
over a large range of scales with high sensitivity.
The array consists of 13 closely packed elements, each of 20cm diameter,
in a configuration which fills roughly half of the aperture area with a
3-fold symmetry.
Each element of the array is a wide-angle corrugated horn with a collimating
lens.  DASI uses cooled HEMT amplifiers running between 26-36GHz with a
noise temperature of $<15$K.  The signal is filtered into ten 1GHz channels.
Details of the layout of the DASI horns is given below.

The outline of this paper is as follows.  In \S\ref{sec:formalism} we adapt
the formalism of \cite{BigPaper} to deal with real and imaginary parts of
the visibilities independently, and show explicitly how the formalism
automatically imposes the constraint that the sky temperature is real.
With the issues defined, we describe the configuration of DASI in
\S\ref{sec:config}.
In \S\ref{sec:osf}, \ref{sec:image} we analyze mock data appropriate to single
fields from DASI, showing how to construct an optimal basis for likelihood
analysis.
We specifically address the question of optimal sampling on the sky, which
was omitted from our previous work.
This leads naturally to a discussion of Wiener filtered
(Bunn, Hoffman \& Silk~\cite{Wiener}) map making and we reformulate the
strategy for imaging the sky in this basis, pointing out the complementarity
between NASAs {\sl MAP\/} satellite\footnote{http://map.gsfc.nasa.gov/} and
DASI at 30GHz.
We discuss estimates of the angular power spectrum which are easy to
implement for uncorrelated pointing of DASI in \S\ref{sec:lucy} and show
that the finite field of view does not hamper our ability to reconstruct
the angular power spectrum.
Finally in \S\ref{sec:radical} we discuss multi-frequency observations and
present our power spectrum results in the context of ``radical compression''
(Bond, Jaffe \& Knox~\cite{BonJafKno}).

\section{Formalism} \label{sec:formalism}

The reader is referred to our earlier paper (\cite{BigPaper}) for a detailed
discussion of how to formulate the data-analysis problem with an
interferometer, plus references to earlier work.  We briefly review the
major elements here.

Under the assumption of a narrow frequency band and a distant source, the
datum measured by an interferometer is proportional to the Fourier Transform
of the observed intensity on the sky, i.e.~the sky intensity multiplied by
the instrument beam.
We label the ``primary'' beam of the telescope by $A({\bf x})$, with
${\bf x}$ a 2D vector lying in the plane\footnote{Since the field of view of
the currently operational or planned instruments is small ($\la 5^\circ$) the
sky can be approximated as flat with excellent accuracy.} of the sky.
Every pair of telescopes in the interferometer measures a visibility at a
given point in the Fourier Plane, called the $u-v$ plane,
\begin{equation}
{\cal V}({\bf u}) \propto \int d{\bf x}\ A({\bf x}) \Delta T({\bf x})
	e^{2\pi i{\bf u}\cdot{\bf x}}
\label{eqn:visdef}
\end{equation}
where $\Delta T$ is the temperature (fluctuation) on the sky and ${\bf u}$
is the variable conjugate to ${\bf x}$, with dimensions of inverse angle
measured in wavelengths.
The (omitted) proportionality constant, $\partial B_\nu/\partial T$ where
$B_\nu$ is the Planck function, converts from temperature to intensity.
The spacing of the horns and the position of the beam on the sky determine
which value of ${\bf u}$ will be measured by a pair of antennae in any one
integration.
The size of the primary beam determines the amount of sky that is viewed,
and hence the size of the ``map'', while the maximum spacing determines the
resolution.

The 2-point function of the observed visibilities is the convolution of the
sky power spectrum, $S({\bf u},{\bf v})$, with the Fourier Transforms of the
primary beams.  If our theory is rotationally invariant the power spectrum
is diagonal, $S({\bf u},{\bf v})=S(u)\delta({\bf u}-{\bf v})$ and on small
scales (\cite{BigPaper})
\begin{equation}
u^2 S(u)\simeq \left. {\ell(\ell+1)\over (2\pi)^2}\ 
  C_\ell \right|_{\ell=2\pi u} \qquad {\rm for}\ u\ga10 \quad.
\label{eqn:suapprox}
\end{equation}
where the dimensionless $C_\ell$ are the usual multipole moments of the CMB
anisotropy spectrum and $\ell\sim\theta^{-1}$ is the multipole index.
This approximation works at the few percent level for a standard Cold Dark
Matter model when $u\ga10$ or $\ell\ga60$.

\begin{figure}
\begin{center}
\leavevmode
\epsfysize=6cm \epsfbox{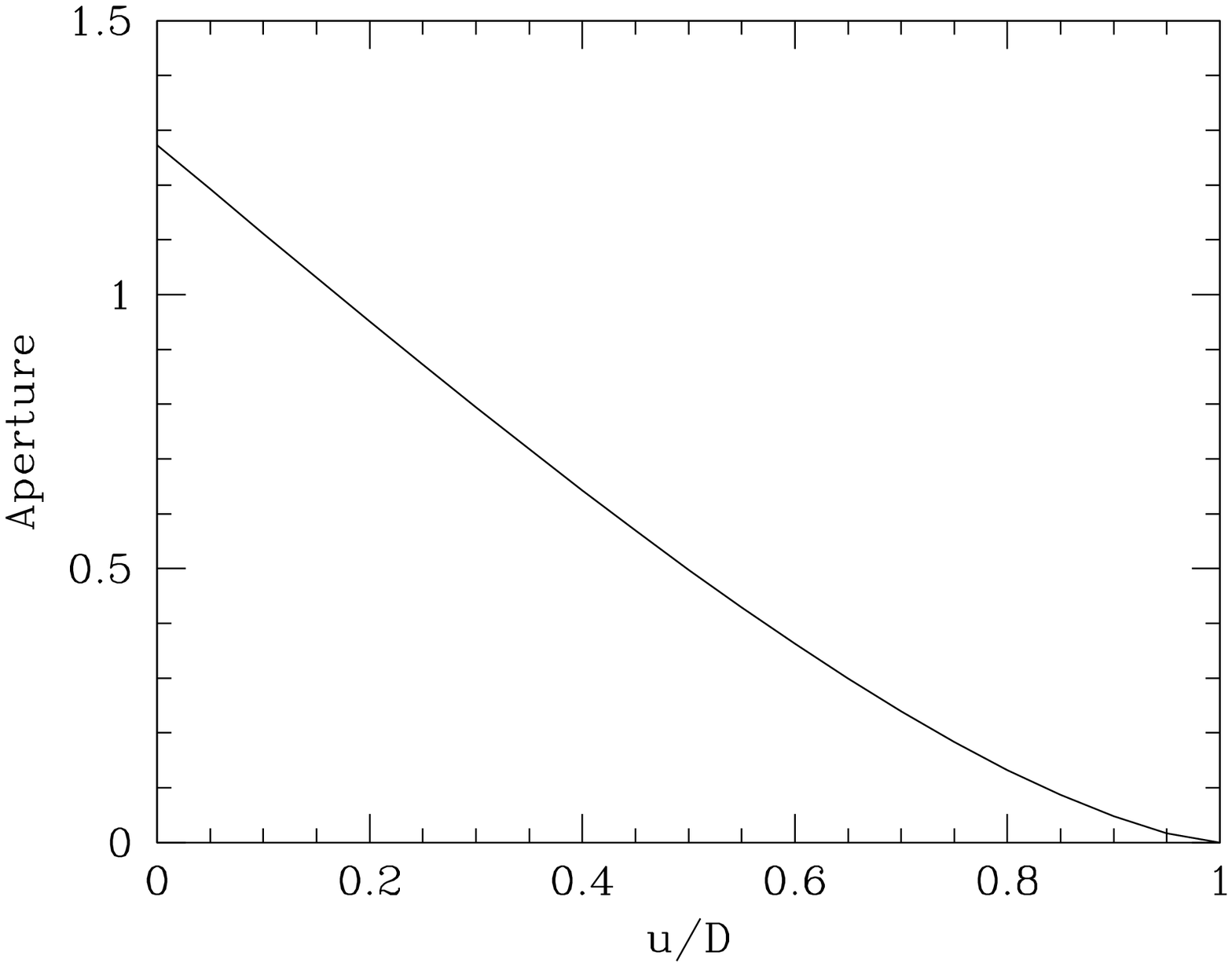}
\epsfysize=6cm \epsfbox{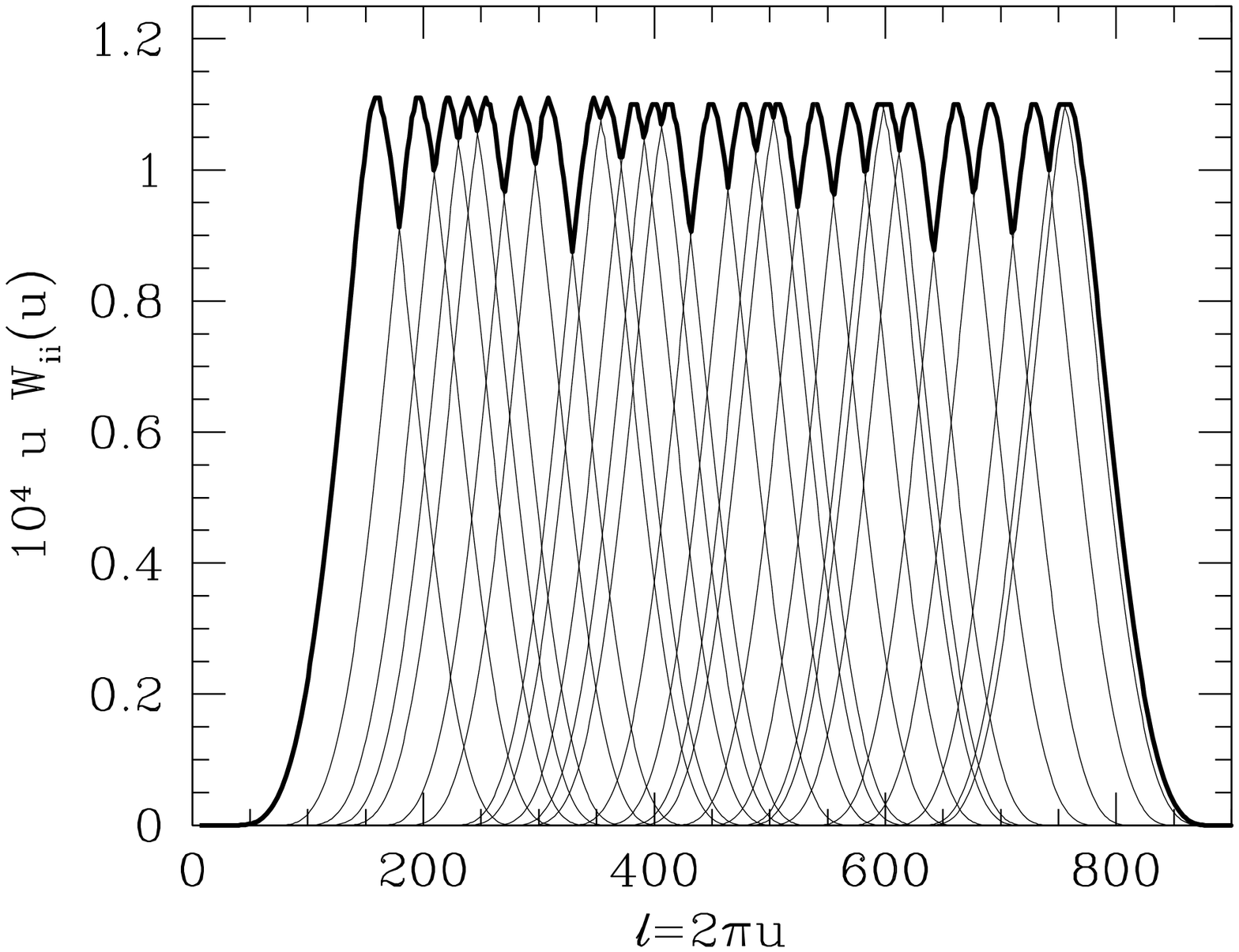}
\end{center}
\caption{(a) The aperture function $\widetilde{A}$ described in the text
(Eq.~\ref{eqn:tildeA}).  For DASI, $D=20$.
(b) The window functions for the 26 baselines of DASI.}
\label{fig:win}
\end{figure}

The Fourier Transform of the primary beam\footnote{Throughout we will use
a tilde to represent the Fourier Transform of a quantity.} is the
auto-correlation of the Fourier Transform of the point response, $g$, of the
receiver to an electric field,
$\widetilde{A}(u)=\widetilde{g}\star\widetilde{g}(u)$ and
\begin{equation}
A({\bf x})=\int d{\bf u}\ \widetilde{A}({\bf u})e^{-2\pi i{\bf u}\cdot{\bf x}}
\quad ,
\end{equation}
Due to the finite aperture $\widetilde{A}$ has compact support.
In order to obtain a simple estimate of our window function it is a
reasonable first approximation to take $\widetilde{A}$ equal to the
auto-correlation of a pill-box of radius $D/2$ where $D$ is the diameter of
the dish in units of the observing wavelength.  Specifically
\begin{equation}
\widetilde{A}({\bf u}) =
  {2\widetilde{A}_*\over\pi} \left[ \arccos{u\over D} -
  {u\sqrt{D^2-u^2}\over D^2} \right]
\label{eqn:tildeA}
\end{equation}
if $u\le D$ and zero otherwise.
If we require $A(0)=1$ then this must integrate to unit area, so
$\widetilde{A}_*^{-1}=\pi(D/2)^2$, or the area of the dish.
We show $\widetilde{A}(u)$ in Fig.~\ref{fig:win}a.
For now we shall treat a single frequency.  Obviously for a fixed physical
dish the $\widetilde{A}$ will be slightly different for different wavelengths.
We return to this complication in \S\ref{sec:radical}.

In \cite{BigPaper} we presented the formalism in terms of complex visibility
data.  However it is easier practically to implement the analysis in terms of
the real and imaginary parts of the visibility.
We write these as ${\cal V}_j\equiv V_j^{R}+iV_j^{I}$.
The cosmological contribution to the real and imaginary components is
uncorrelated $\left\langle V_i^RV_j^I\right\rangle=0$.
Assemble $V^R$ and $V^I$ into a vector consisting of first the real and
then the imaginary parts -- the signal correlation matrix of this vector
takes block-diagonal form.
Further the cosmological contribution obeys
$\left\langle V_i^RV_j^R\right\rangle = \pm
 \left\langle V_i^IV_j^I\right\rangle$.
It is straightforward to show that the cosmological contribution is
proportional to
\begin{equation}
  {1\over 2}\int d{\bf v} \ S(v) \widetilde{A}({\bf u}_i-{\bf v})
  \left[ \widetilde{A}({\bf u}_j-{\bf v}) \pm
         \widetilde{A}({\bf u}_j+{\bf v}) \right]
\end{equation}
where the $\pm$ refer to the real and imaginary parts respectively.
(We have dropped the normalization factor which converts temperature to flux.)
Note that if ${\bf u}_i=-{\bf u}_j$ the visibilities are completely
(anti-)correlated, as would be expected given that ${\cal V}({\bf u})$ is the
Fourier Transform of a real field: ${\cal V}^{*}({\bf u})={\cal V}(-{\bf u})$.
The factor of ${1\over 2}$ out front reflects the fact that the full variance
$C^V_{ij}\equiv \left\langle {\cal V}^{*}_i{\cal V}_j\right\rangle$ is the sum
of the real and imaginary components.

In the case where all correlated signal is celestial, the correlation function
of the noise in each visibility is
diagonal with
\begin{equation}
C^N_{ij}=\left( {2k_BT_{\rm sys}\over\ {\eta_A A_D} } \right)^2
         {1\over \Delta_\nu t_a n_b}\ \delta_{ij} \quad .
\label{eqn:cndef}
\end{equation}
If the noise in the real and imaginary components is uncorrelated, then each
makes up half of this variance.
Here $k_B$ is Boltzmann's constant, $T_{\rm sys}$ is the system noise
temperature, $\eta_A$ is the aperture efficiency, $A_D$ is the physical area
of a dish (not to be confused with $A({\bf x})$), $n_b$ is the number of
baselines\footnote{The number of baselines formed by $n_r$ receivers is
$n_b=n_r(n_r-1)/2$.} corresponding to a given separation of antennae,
$\Delta_\nu$ is the bandwidth and $t_a$ is the observing time.
Typical values for DASI are $T_{\rm sys}=20$K, $\eta_A\sim0.8$, dishes
of diameter 20cm, $n_b=3$ and $\Delta_\nu=10$GHz (in $10\times1$GHz channels).

\begin{figure}
\begin{center}
\leavevmode
\epsfysize=6cm \epsfbox{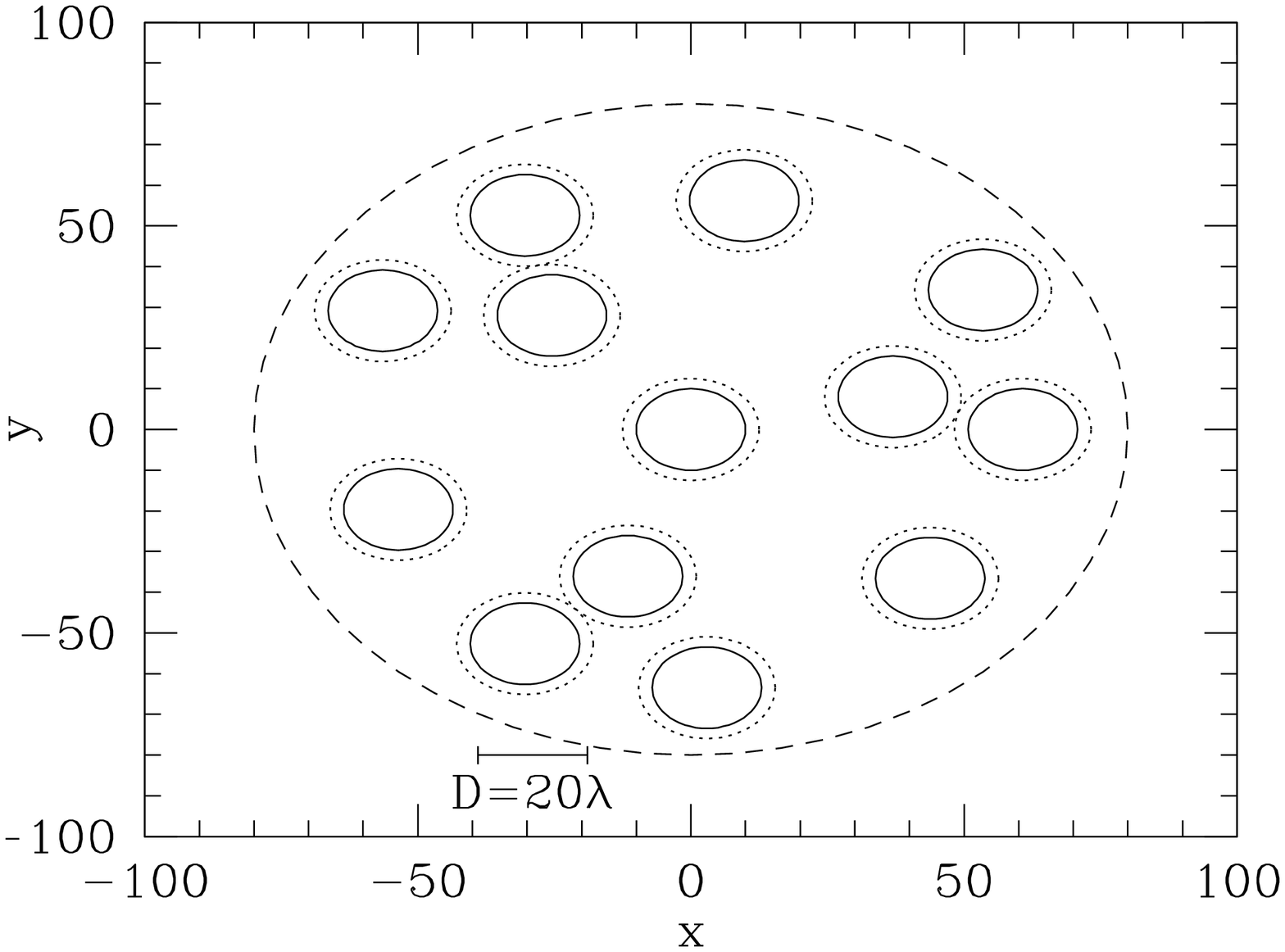}
\epsfysize=6cm \epsfbox{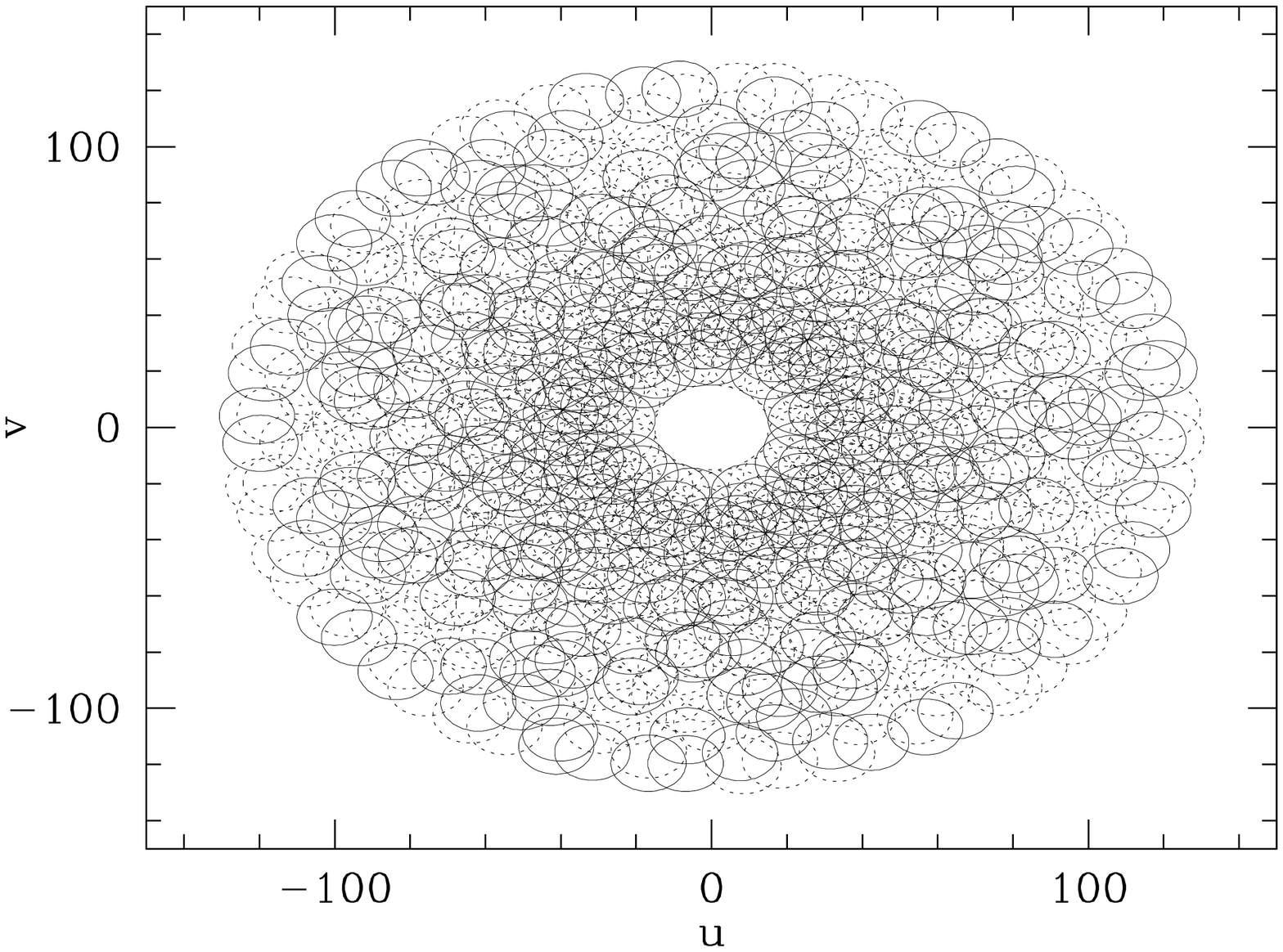}
\end{center}
\caption{(a) The positions of the dishes for DASI.  Each dish has an
effective optical diameter $D=20\lambda$ with $\lambda=1$cm (solid lines).
The physical size of the dishes is 25cm (dotted lines) and the base plate
for DASI is 1.6m in diameter (dashed line).
(b) The positions at which visibilities will be measured in 1 ``stare'' of
DASI.  The circles show apertures of radius 10 (roughly the half power point
of the beam profile).
Each visibility consists of an independent real and imaginary part.
Solid lines indicate where data will be taken, the dotted lines indicate
points constrained by symmetry of the transform.}
\label{fig:vis}
\end{figure}

We show in Fig.~\ref{fig:vis} the positions at which DASI will measure
visibilities.  With $13$ horns there are $78$ baselines.  Because of the
3-fold symmetry of the instrument each pointing samples $V$ at $26$ different
$|{\bf u}|$.  For each ``stare'' DASI is rotated about an axis perpendicular
to the baseplate to fill half of the $u-v$ plane as shown in
Fig.~\ref{fig:vis}.  The other half of the $u-v$ plane is constrained by
the symmetry ${\cal V}^{*}({\bf u})={\cal V}(-{\bf u})$.

\section{The DASI configuration} \label{sec:config}

The layout of the horns for DASI is shown in Fig.~\ref{fig:vis}a.  This
configuration has a 3-fold symmetry about the central horn.
The positions of 4 of the non-central horns are arbitrary, up to a global
rotation, with the remaining configuration being determined by symmetry.
The configuration shown represents an optimal configuration, within the
physical constraints, for the purposes of measuring CMB anisotropy.

As described in the last section, the distance between each pair of horns
represents a baseline at which a visibility can be measured.  Each visibility
probes a range of angular scales centered at $\ell=2\pi u$ where $u$ is
the baseline in units of the wavelength.  The sensitivity as a function of
$\ell$ is given by the window function (see Eq.~\ref{eqn:wadef}).
Since the width of the window function is determined by the size of the
apertures, the optimal coverage for the purpose of CMB anisotropy is that
configuration which spans the largest range of baselines with the most
overlap between neigbouring window functions.
In one dimension such optimal configurations are known as
Golomb\footnote{e.g.~http://members.aol.com/golomb20/} rulers
(Dewdney~\cite{Golomb}), and a well known procedure exists for finding them.

Finding an optimal solution in two dimensions, within the physical
constraints, cannot be done analytically.  We have optimized the configuration
numerically.
We have searched the 7 dimensional parameter space of horn positions (an
$x$ and $y$ position for each of 4 horns, minus one overall rotation) for
the configuration which minimized the maximum separation between baseline
distances, while at the same time covering the largest range of angular
scales.  Trial starting positions were determined by a simple Monte Carlo
search of the parameter space.  From each of these positions a
multi-dimensional minimization was started.
Two additional constraints were imposed upon the allowed solutions: no two
horns could come closer than 25cm, the physical size of the horn plus
surrounding ``lip'', and all horns etc had to lie completely within the 1.6m
diameter base plate (leading to a maximal radius of $63.5$cm).
Our ``optimal'' solution is given in Table~\ref{tab:config}.
The baselines run from 25cm to 120cm with the largest gap between
baseline distances being 6.4cm.

\section{S/N Eigenmodes or Optimal Subspace Filtering} \label{sec:osf}

\begin{figure}
\begin{center}
\leavevmode
\epsfysize=6cm \epsfbox{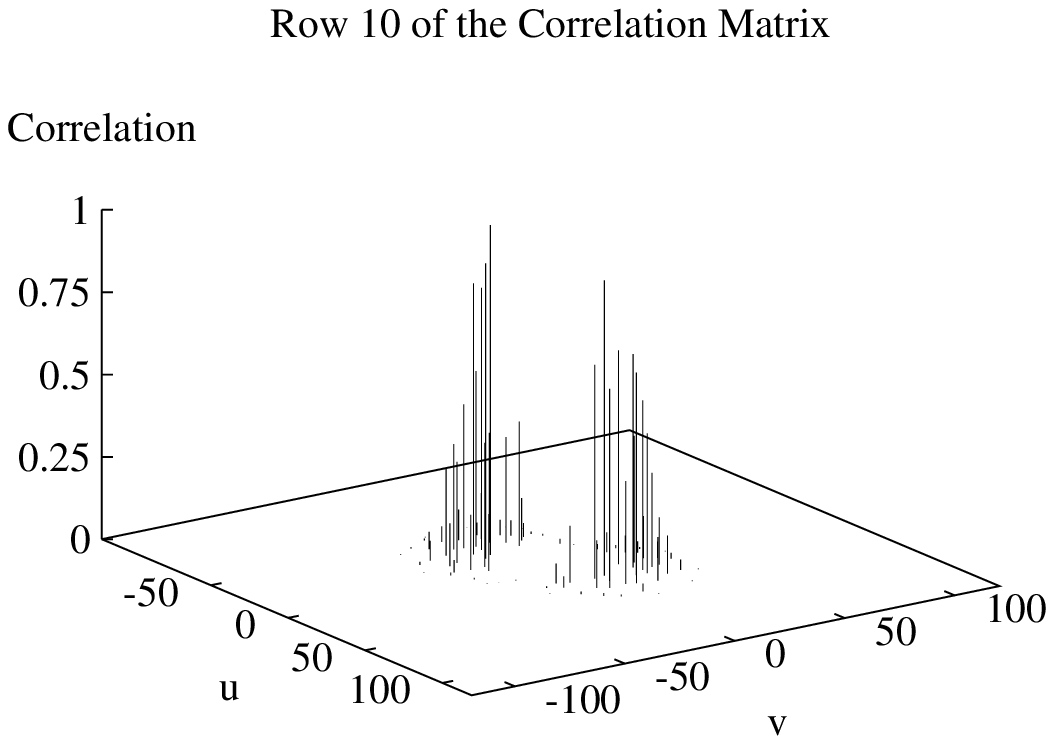}
\epsfysize=6cm \epsfbox{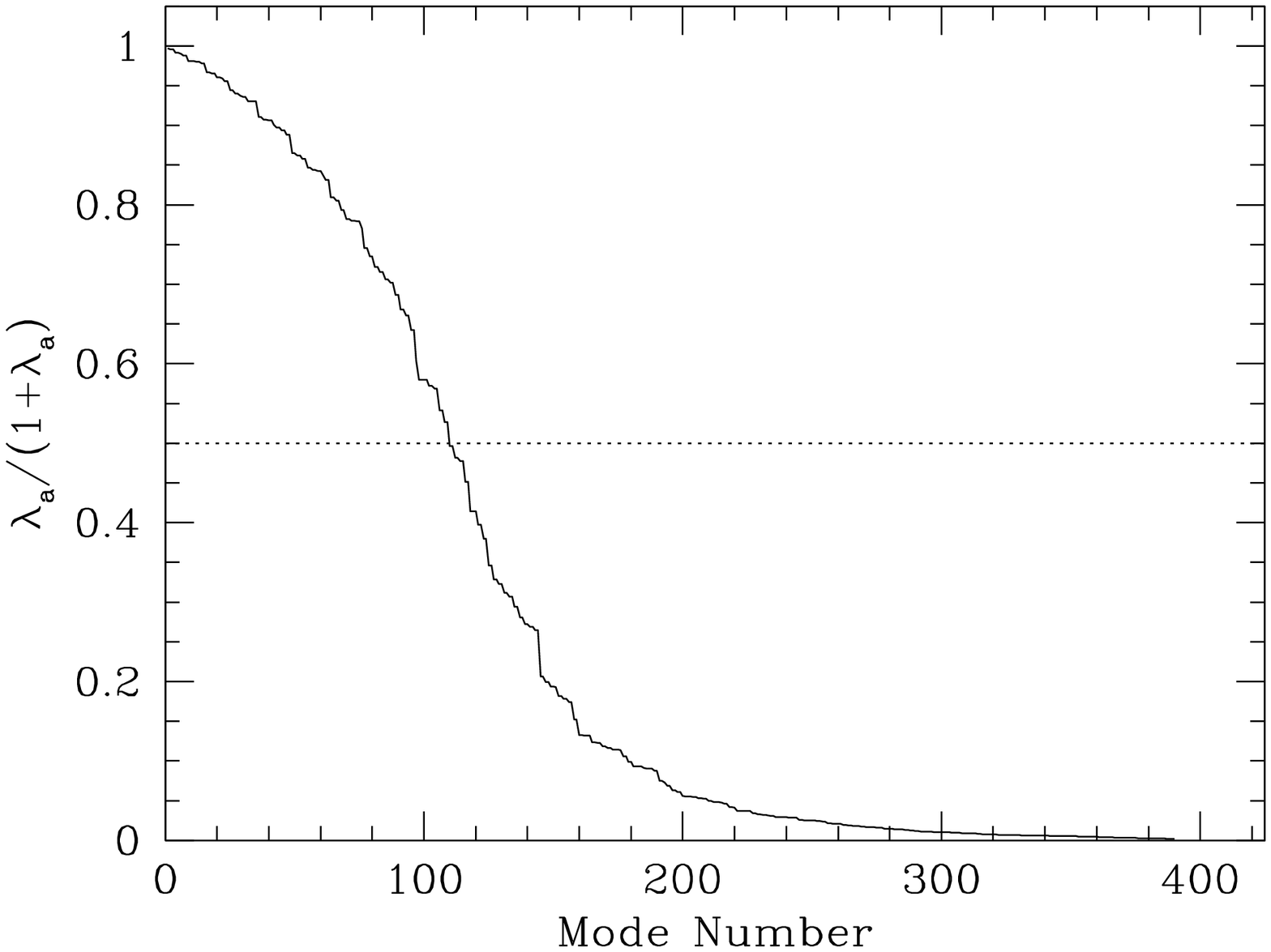}
\end{center}
\caption{(a) One row of the signal correlation matrix, plotted in the $u-v$
plane.
(b) the signal-to-noise eigenvalues $\lambda_a$ for the {\it real\/} part
of the visibilities.  For every real $\lambda_a$ there is an independent
imaginary mode with the same signal-to-noise.
All calculations assume a scale-invariant COBE normalized spectrum.}
\label{fig:eigen}
\end{figure}

A quick glance at Fig.~\ref{fig:vis}b shows that the signal in most of the
visibilities will be highly correlated, i.e.~the apertures have a large
overlap.  This is shown quantitatively in Fig.~\ref{fig:eigen}a where we plot
one row of the signal correlation matrix.
Given a ``trial'' theory we can perform a change of basis to remove these
correlations (see \S6.2 of \cite{BigPaper}, or Tegmark et al.~\cite{THSVZ}
for a review of this method).
The input theory can be considered as a prior in the context of Bayesian
analysis, or could be iterated to match the data if so desired.  For our
purposes all that will matter is that the signal variance is independent of
$\widehat{u}$ and decreases with $|{\bf u}|$.  For concreteness we will
take $u^2 S(u)=$constant, normalized to the COBE 4-year data
(Bennett et al.~\cite{4year}).  For this choice the cosmological signal is
approximately equal to the noise in each of the highest $|{\bf u}|$ bins in
Fig.~\ref{fig:vis}b.

Let us consider only the real parts of the visibilities for now, the imaginary
parts are dealt with in an analogous manner.
Take these to lie in the upper-left block of the block-diagonal correlation
matrix.  Denote the noise in the real part of each visibility $\sigma_j$.
The eigenvalues of the matrix $C^{RR}_{ij}/\sigma_i\sigma_j$ measure the
signal-to-noise in {\it independent\/} linear combinations of the visibilities.
The independent linear combinations can be written
$\nu_a = \sum_i (V_i/\sigma_i) \Psi_{ia}$ where $\Psi_{ia}$ is the $a$th
eigenvector of $C^{RR}_{ij}/\sigma_i\sigma_j$.  Then
$\left\langle\nu_a\nu_b\right\rangle=(\lambda_a+1)\delta_{ab}$ where the
$\lambda_a$ are the signal-to-noise eigenvalues and the 1 represents the
noise contribution (which is the unit matrix in this basis).

The number of modes $\nu_a$ with $\lambda_a>1$ is a quantitative indication
of how much signal is being measured by the experiment.  We show the
eigenspectrum of the {\it real\/} parts of the visibilities for 1 day of
observation in Fig.~\ref{fig:eigen}b.  There is in addition an eigenvalue of
the same size for each imaginary component.
Due to the overlapping windows in the $u-v$ plane, $\la 25\%$ of the modes
contain good cosmological information.  The rest are redundant, containing
mostly noise.
Note however that in 1 day of observation DASI measures nearly 200 high
signal-to-noise eigenvectors!
We have tested that reducing the number of orientations of the instrument,
i.e.~the oversampling in $\theta_u$, does not significantly reduce the
number of high signal-to-noise eigenvectors until the apertures in the
outermost circle are just touching, at which point there are 160 modes
with $S/N\ga 1$.

The $\lambda_a$ are the variance of the $\nu_a$ and can be predicted given
a theoretical power spectrum
\begin{equation}
  \lambda_a = {1\over 2} \int vdv\ S(v) W_a(v)
\end{equation}
with the window function
\begin{equation}
W_a(|{\bf v}|)= \sum_{ij} {\Psi_{ia}\over\sigma_i} {\Psi_{ja}\over\sigma_j}
  \ \int d\theta_v\ \widetilde{A}({\bf u}_i-{\bf v})
  \left[ \widetilde{A}({\bf u}_j-{\bf v}) \pm
         \widetilde{A}({\bf u}_j+{\bf v}) \right]
\label{eqn:wadef}
\end{equation}
and as such could form a basis for ``radical compression''.  We shall return
to this issue in \S\ref{sec:radical}.

\section{Imaging the Sky} \label{sec:image}

DASI provides coverage over a significant part of the $u-v$ plane, and as
such is able in principle to perform high resolution imaging of the sky.
The high signal to noise of the DASI instrument and the large dynamic range
in angular scale however make imaging a computationally challenging task.
In this section we present two approaches which we have tried with mixed
success on simulated data.

We discussed the Wiener filtering formalism for producing an image from the
visibility data in \cite{BigPaper}.  Here we note that in the $\nu_a$ basis
it is straightforward to construct the Wiener filtered sky map
(Bunn, Hoffman \& Silk~\cite{Wiener};
\cite{BigPaper} and references therein).
Recall that for Wiener filtering the sky temperature is approximated by
\begin{equation}
  T^{WF}({\bf x}_\alpha) = C^{T}_{\alpha\beta} W_{\beta j}
     \left[ C^V+C^N \right]_{jk}^{-1}\quad V_k
\end{equation}
where $C^T$ is the real-space temperature correlation function,
\begin{equation}
W_{\beta j}= A\left({\bf x}_\beta\right)
  \left\{ {\cos\atop\sin} \right\}
  \left[ 2\pi\,i\, {\bf u}_j\cdot{\bf x}_\beta \right]
\end{equation}
we choose $\cos$ for the real components and $\sin$ for the
imaginary components, and $C^V$ and $C^N$ are the visibility signal and
noise correlation matrices.
We have written the expression in this mixed way to avoid needing to invert a
matrix of size $N_{\rm pix}\times N_{\rm vis}$ where $N_{\rm pix}$ is the
number of pixels in the map being reconstructed, but conceptually the
$C^T W$ term can be replaced with $W^{-1}C^V$.
Then in the $\nu_a$ basis one simply replaces $\nu_a$ with
$\lambda_a(\lambda_a+1)^{-1}\nu_a$ which down-weights modes with low
signal-to-noise.  Specifically
\begin{equation}
  T^{WF}({\bf x}_{\beta} ) = \sum_a
  M_{\beta a}^{-1}\ {\lambda_a\over\lambda_a+1}\, \nu_a
\end{equation}
where
\begin{equation}
M_{\beta a}=\sum_j {\Psi_{ja}\over \sigma_j}
  \ A\left({\bf x}_\beta\right)
  \left\{ {\cos\atop\sin} \right\}
  \left[ 2\pi\,i\, {\bf u}_j\cdot{\bf x}_\beta \right]
\end{equation}
where we choose $\cos$ for the real components and $\sin$ for the
imaginary components.
The ratio $\lambda_a(\lambda_a+1)^{-1}$ is plotted in Fig.~\ref{fig:eigen}b,
where we can see that $\la 25\%$ of the modes will contribute significantly
to the final map.
The expected variance in the final map is also easily computed:
\begin{equation}
  \left\langle T^{WF}_\rho T^{WF}_\sigma\right\rangle=
 \sum_a M_{\rho a}^{-1} {\lambda_a^2\over \lambda_a+1} M_{a\sigma}^{-1}
\end{equation}
which approaches $\left\langle T^2\right\rangle$ as $\lambda_a\to\infty$.
Note that the Wiener filter is {\it not\/} power preserving, so the maps
should not be used for power spectrum estimation.

By reducing the total number of modes that need to be kept in the summation
the $\nu_a$ basis can speed up calculation of the Wiener filtered map.  This
is an advantage, but even with this speed up we found that producing a high
resolution map of even a single field on the sky is very computationally
expensive due to the large number of matrix multiplications involved in
computing $T^{WF}$.
An alternative to Wiener filtering, which is very similar in the high
signal-to-noise regime in which we are working, is the minimum variance
estimator for $T({\bf x})$.
The number of operations required to produce the minimum variance and
Wiener filtered maps are comparable, and both tend to be very slow.  We
are currently investigating faster approximate methods of image making.

The formalism above can in principle be used for making maps at each
frequency, or one can make a map which combines the frequencies in such a
way as to isolate the CMB (or foreground) signal.
While the formalism looks more complex, it is in fact easy to implement
computationally.
The techniques are well known (see \S\ref{sec:radical}):
we imagine that at each point in the $u-v$ plane our visibilities form a
vector $\vec{V}$ whose components are the different frequencies.
We can expand this vector in terms of the physical components we wish to
consider, e.g.,
\begin{equation}
  \vec{V} = \theta_{\rm CMB} \vec{V}^{\rm CMB} + \theta_{ff} \vec{V}^{ff}
\end{equation}
In the absence of noise it is easy to solve for $\theta_{\rm CMB}$ as
\begin{equation}
  \theta_{\rm CMB} =
  {\vec{V}_\perp\cdot\vec{V}\over\vec{V}_\perp\cdot\vec{V}_{\rm CMB}}
  \qquad {\rm with}\qquad \vec{V}_\perp\cdot\vec{V}_{ff}=0
\end{equation}
If we write $\theta_{\rm CMB}=\sum_a c_a V_a$ then we should replace
$\widetilde{A}$ by $\sum_a c_a \widetilde{A}_a$ throughout.
In the presence of noise we may define $c_a$ by a minimum variance estimator
as above:
\begin{equation}
  c_a = \sum_{A} \left( \sum_{bc} V_b^{\rm CMB} N^{-1}_{bc} V_c^{A}\right)^{-1}
        \sum_d V_d^{A} N^{-1}_{da}
\end{equation}
where $N_{ab}$ is the channel noise matrix and capital Roman letters indicate
the component being considered (see \S\ref{sec:radical}).

\begin{figure}
\begin{center}
\leavevmode
\epsfysize=6cm \epsfbox{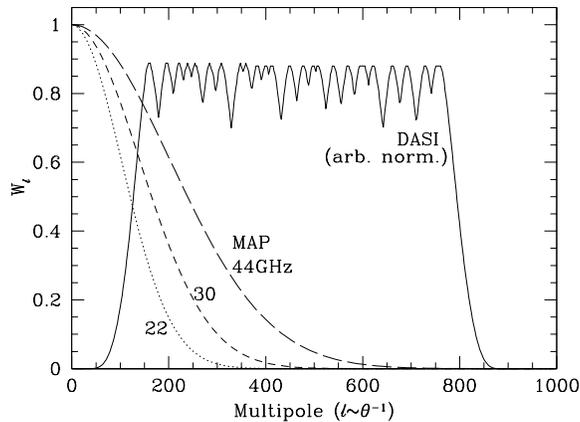}
\end{center}
\caption{The window functions for {\sl MAP\/} and DASI, showing the
complementarity in angular scale of the two instruments at low frequency.
The {\sl MAP\/} data will provide large-scale information over the whole
sky which can be used when making maps with DASI, which is sensitive to
the small angular scales which {\sl MAP\/} cannot resolve.}
\label{fig:MAPwindow}
\end{figure}

We end our discussion of imaging with an observation about the complementarity
between the {\sl MAP\/} satellite and DASI.
In making maps of the sky with DASI, the largest source of error is the
missing long wavelength modes which are filtered out by the DASI primary
beam.  The long wavelength power can be included in the map by simultaneously
fitting another data set (e.g.~\cite{BigPaper}).
In this particular case the data from the {\sl MAP\/} satellite at the same
frequency is the obvious choice since {\sl MAP\/} will have full sky coverage.
We show the window functions for {\sl MAP\/} at the relevant frequencies,
along with the envelope for DASI in Fig.~\ref{fig:MAPwindow}.

\section{Power Spectrum Estimation} \label{sec:lucy}

While the $\nu_a$ are the natural basis from the point of view of
signal-to-noise and Wiener filtering, and can dramatically improve likelihood
function evaluation, they are not necessarily the quantities of greatest
physical interest for power spectrum estimation.
In \cite{BigPaper} we discussed estimating a series of bandpowers using the
quadratic estimator of
(Bond, Jaffe \& Knox~\cite{BonJafKno}, Tegmark~\cite{Max}).
Here we present a simpler approach more appropriate to unmosaiced fields.
In this presentation we shall assume we are dealing with single frequencies;
the case of multiple frequencies is dealt with in \S\ref{sec:radical}.

In CMB anisotropy observations with single dish experiments much has been
written about optimal ways to estimate the power spectrum.  The principle
reason is that care must be taken in weighting the data to ensure that no
sharp cut-offs in real space are introduced.
These lead to ringing in Fourier space and delocalize the window function
(e.g.~discussion in Tegmark~\cite{Teg}).
For the interferometer no such problem arises -- each visibility samples a
compact region in ${\bf u}$.
For each $V_i$, the square is a noise biased estimate of the power
spectrum convolved with a window function.
If we define $s_i \equiv 2\left(V_i^2 - N_{ii}\right)$ with $N_{ii}$ the noise
variance in (the component of the) visibility $i$, then $s_i$ is an unbiased
estimator for $C^V_{ii}$:
\begin{equation}
  \left\langle s_i\right\rangle = \int u\,du\ S(u) W_{ii}(u)
\label{eqn:suest}
\end{equation}
where $W_{ii}(u)$ is the window function in analogy to Eq.~(\ref{eqn:wadef}).
We have included the factor of 2 in the definition of $s_i$ to account for the
fact that each component $V_i$ of the visibility contributes half of the
total variance of $|{\cal V}_i|^2$.
Take a weighted average of the $s_i$: ${\cal S}_A \equiv E_{Ai} s_i$.
The simplest weighting is to sum all of the $s_i$ with $|{\bf u}_i|\simeq u$
and we shall use that below.
Under the assumption that the visibilities are Gaussian, the error matrix for
the estimates ${\cal S}_A$ is
\begin{equation}
  \left\langle \delta{\cal S}_A\ \delta{\cal S}_B \right\rangle =
   2\sum_{ij} E_{Ai}\left( C^V_{ij}+C^N_{ij} \right)^2 E_{jB}
  \quad .
\end{equation}
Note that for $N$ uncorrelated visibilities (real and imaginary parts)
contributing to ${\cal S}_A$ this gives
$\delta{\cal S}_A/{\cal S}_A=N^{-1/2}(1+{\rm noise}/{\rm signal})$ as
expected.

For DASI in the configuration shown in Fig.~\ref{fig:vis}b one can construct
26 different estimates ${\cal S}_A$, which will however be quite correlated.
We show the $13\times13$ correlation matrix for every {\it second\/} estimate
in Table~\ref{tab:corr} along with the expected error on each determination
(the diagonal elements).  As independent pointings are included in the
analysis the error bar on each ${\cal S}_A$ decreases as $N^{-1/2}$, but the
correlations remain the same.  Without mosaicing (\cite{BigPaper}) the
resolution in $\ell$ is restricted to $\sim 2\pi D$, thus the individual
determinations are required to be highly correlated.
The error on the highest $\ell$ bins is still dominated by the small number
of independent samples, in this case 16.  Increasing the oversampling in the
angular direction (and the observing time) can reduce this error to about
23\%, hardly worth the extra time compared to including a different pointing.

The ${\cal S}_A$ are estimates of the power spectrum {\it convolved with
the window function\/} $W_A(u)$.  Given a theory it is straightforward to
compare to the data once the window functions are known, and we discuss this
in \S\ref{sec:radical}.
However one could ask whether (or how much) cosmological information has been
lost due to the convolution or if the information is still present in the
correlated visibilities.
One way of answering this question is to attempt to perform an approximate
(theory independent) deconvolution.

\begin{figure}
\begin{center}
\leavevmode
\epsfysize=6cm \epsfbox{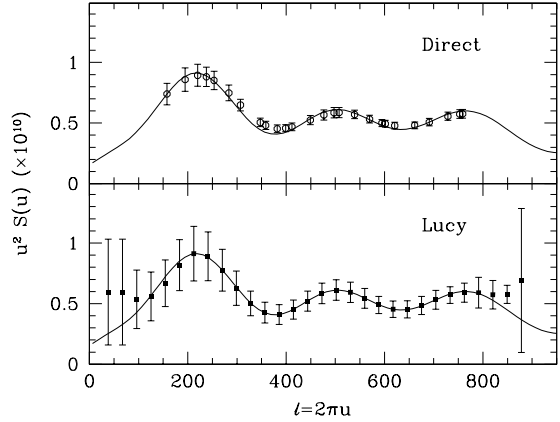}
\end{center}
\caption{Estimates of the power spectrum obtained from simulated data
from 16 independent DASI fields, each observed for 1 day.
(Upper) The open circles indicate estimates obtained by assuming
$S(u)=$constant through the window function.
(Lower) The solid squares are from Lucy's method.  The points are highly
correlated (see text).}
\label{fig:deconvolve}
\end{figure}

The simplest (``direct'') method is to assume that $S(u)=$constant through
the window, so that $S(u)\approx {\cal S}_A/\int u\,du\ W_A(u)$ with
$W_A(u)=\sum_i E_{Ai} W_{ii}(u)$.
(Alternatively one could assume that $u^2S(u)=$constant, which leads to a
similar expression.)
For the window function of Eq.~(\ref{eqn:tildeA}) we have
$\int d^2u\, \widetilde{A}^2(u)\simeq 0.585 D^{-2}$.
As we show in Fig.~\ref{fig:deconvolve}, this technique works surprisingly
well for DASI, even though the FWHM of each window is $\Delta\ell\sim100$,
over which scales we can expect power spectra to change significantly.
For a CDM power spectrum for example, the approximation above induces a
systematic 10\% error (at worst) due to the window function ``washing out''
the peak structure.

To avoid this problem, we can attempt to perform the deconvolution by an
iterative procedure.
Recall that we are trying to constrain the power spectrum, while our
measurements are the spectrum convolved with a positive semi-definite
kernel -- the window function.
In this situation Lucy's method (Lucy~\cite{Lucy}) can be used.
We have implemented Lucy's algorithm, following
Baugh \& Efstathiou~(\cite{BauEfs}), including a coupling between different
bins and a regularization of the iteration.  As they found, the final result
is not sensitive to the mechanism chosen.

To briefly re-cap the method: we think about the deconvolution problem,
following Lucy~(\cite{Lucy}), first in terms of probability distributions.
If we denote by $p(x)$ the probability of measuring a quantity $x$ and
$p(y|x)$ the conditional probability of measuring $y$ given that $x$ is
true then
\begin{equation}
  p(y) = \int p(y|x) p(x) dx
\label{eqn:probeqn}
\end{equation}
which is a convolution integral.
We wish to estimate $p(x)$ given observations $p^{\rm obs}(y)$.
We start the $r$th iteration with an estimate $p^r(x)$ of $p(x)$ and
predict $p^r(y)$ using Eq.~(\ref{eqn:probeqn}) assuming $p(y|x)$ is known.
Writing the inverse of Eq.~(\ref{eqn:probeqn}), for $p(x)$, using the observed
$p^{\rm obs}(y)$ and rewriting $p(x|y)p(y)=p(y|x)p(x)$ leads us to the
iterative expression for $p^{r+1}(x)$:
\begin{equation}
  p^{r+1}(x) = p^r(x)
  \ {\int \left(p^{\rm obs}(y)/p^r(y)\right) p(y|x) dy \over
     \int p(y|x) dy }
\end{equation}
where the denominator is unity.
The iterative method we use takes this expression over with the replacements
$p(y)\to{\cal S}_A$, $p(x)\to f(u)\equiv u^2 S(u)$ and
$p(y|x)\to u^{-1}W_A(u)$.  Approximating the integrals as sums equally spaced
in $u$ we have the iterated pair of equations:
\begin{eqnarray}
 {\cal S}_A^r &=& \sum_a f^r(u_a)\, u_a^{-1}W_A(u_a)\ \Delta u \\
  f^{r+1}(u_a) &\equiv& f^r(u_a)
  \ {\sum_i \left({\cal S}^{\rm obs}_A / {\cal S}^r_A\right)
   u_a^{-1}W_A(u_a) \over \sum_i u_a^{-1}W_A(u_a) }
\end{eqnarray}
To make the iteration converge more stably we in fact replace only a
fraction $\epsilon$ of $f^r$ with $f^{r+1}$ on each step, and average
the $f^r(u_a)$ using a 2nd order Savitsky-Golay filter of length $(2,2)$.
The final result is insensitive to the details of this procedure, the number
of bins chosen for $u_a$ etc.

We show in Fig.~\ref{fig:deconvolve} results of a deconvolution assuming
that 16 independent patches of sky were observed, each for 1 day.
We can see that the deconvolution works well, with good resolution in $\ell$
and no bias over the scales where DASI has sensitivity.  Thus we expect that
no important cosmological information has been lost by the convolution
procedure.
The points are highly correlated, but we do not show the correlations on
the Figure (all correlations were included in the analysis).  The error bars
are so large because they are the computed from the variance {\it allowing
all other bins to vary freely\/}.

Though the example here was for a single frequency channel, the generalization
to multiple frequencies is straightforward and can in principle allow even
finer sub-band resolution.

Finally we remark that obviously a statistical comparison of a given model
to the data should be done with the ${\cal S}_A$.
The correlations of the ${\cal S}_A$ can be computed for any given theory
(e.g.~Table~\ref{tab:corr}).
This allows a full likelihood analysis to be done, and is the route which
should be taken when comparing DASI data to a specific theory.  This is
what we shall discuss now.

\section{Radical Compression} \label{sec:radical}

To constrain theories using interferometer data is much easier than in the
case of single dish data, thus many of the powerful techniques developed
for the latter case are not needed.
The primary reason for this is that interferometers work directly in the
$\ell$-space of theories.  It is straightforward to perform ``radical
compression'' (Bond, Jaffe \& Knox~\cite{BonJafKno}) of an interferometer
data set and quote a set of bandpowers along with their full noise covariance
matrix and window functions. 
We shall develop this idea briefly in this section.

We now reintroduce the multi-frequency nature of the data set that has been
suppressed during most of this paper.
For DASI we can work with 10$\times$1GHz channels which we shall label with
a greek subscript.
Following Dodelson~(\cite{FDF}; see also White~\cite{BeamComment}) we
imagine that our visibility signal $V_\alpha$ at each point in the $u$-$v$
plane is a sum of contributions with different frequency dependences:
$V_a = \sum_A \theta^A V^A_a$.  Let $A=0$ be the CMB contribution,
whose frequency dependence will be $V^0=(1,1,\cdots)$ for observations at
DASI frequencies.
At each frequency, $a$, and visibility position, $i$, the signal is
the convolution of the sky with an aperture $\widetilde{A}^a_i$ which will
be of the form of Eq.~(\ref{eqn:tildeA}) with the central $u$ and $D$ varying
by the inverse of the observing wavelength.
If each visibility has noise $N_{ab}^{(i)}$ we can estimate the
CMB component $\theta^0$ by minimizing
\begin{equation}
\chi^2 = \sum_{ab} \sum_{AB}
         \left( V_a^{\rm obs} - \theta^A V^A_a \right)
         N_{ab}^{-1}
         \left( V_b^{\rm obs} - \theta^B V^B_b \right)
\end{equation}
where we have suppressed the visibility index $i$.
Solving $d\chi^2/d\theta^A=0$ amounts to taking a linear combination of the
frequency channels $\theta_i^{0}=\sum_a c_a V^{\rm obs}_a$ with
\begin{equation}
c_a = \sum_A \left( \sum_{bc} V_b^0 N_{bc}^{-1} V_c^A \right)^{-1}
      \sum_{d} V_d^A N_{ad}^{-1}
\end{equation}
While formidable this expression reduces to the well known least-squares
weighting in the limit $N_{ab}=\sigma_a^2\delta_{ab}$:
\begin{equation}
  c_a = \sum_{Ab} \left( {V_b^0 V_b^A \over \sigma_b^2} \right)^{-1}
        {V_a^A\over\sigma_a^2}
\end{equation}

Now we simply replace $V_i$ with $\theta_i^0$ and $\widetilde{A}_i$ with
$\sum_a c_a \widetilde{A}_{ia}$ to generalize Eq.~(\ref{eqn:suest}).
The generalized ${\cal S}_A$ can be easily calculated from the data,
and the expectation values and distribution can be calculated for any
theory once the window function and noise properties are given.
Supplying this set would be an ideal way to release the DASI data for the
purposes of model fitting.

\section{Conclusions}

We have shown how one can implement the formalism of \cite{BigPaper} in
the case of a single stare of the DASI instrument.  We have discussed
making maps, filtering the data and reconstructing the anisotropy power
spectrum.  Our results suggest that with $\sim 1$ month of data, DASI
could provide significant constraints on theories of structure formation.

\bigskip
\acknowledgments  
We would like to acknowledge useful conversations with Alex Szalay and
thank Tim Pearson for comments on a draft of this work.

\begin{table}
\begin{center}
\begin{tabular}{cc}
 $x$ (cm) & $y$ (cm) \\
\hline
 $+00.00$ & $+00.00$\\
 $+60.76$ & $+00.00$\\
 $+53.47$ & $+34.25$\\
 $+36.98$ & $+08.04$\\
 $+09.77$ & $+56.21$\\
 $-30.38$ & $+52.62$\\
 $-56.40$ & $+29.18$\\
 $-25.45$ & $+28.01$\\
 $-53.56$ & $-19.64$\\
 $-30.38$ & $-52.62$\\
 $+02.93$ & $-63.43$\\
 $-11.53$ & $-36.05$\\
 $+43.79$ & $-36.57$
\end{tabular}
\end{center}
\caption{The positions of the centers of the horns on the DASI base plate
for the optimal configuration described in \S\ref{sec:config}.}
\label{tab:config}
\end{table}

\begin{table}
\begin{center}
\begin{tabular}{c|ccccccccccccc}
$\ell_i=2\pi u_i$
& 158& 220& 253& 307& 358& 399& 449& 497& 539& 595& 621& 691& 753\\
\hline
 158&100& 36&  9&  0&  0&  0&  0&  0&  0&  0&  0&  0&  0\\
 220& 36&100& 72& 14&  1&  0&  0&  0&  0&  0&  0&  0&  0\\
 253&  9& 72&100& 45&  6&  0&  0&  0&  0&  0&  0&  0&  0\\
 307&  0& 14& 45&100& 45& 11&  0&  0&  0&  0&  0&  0&  0\\
 358&  0&  1&  6& 45&100& 52& 10&  1&  0&  0&  0&  0&  0\\
 399&  0&  0&  0& 11& 52&100& 39&  6&  0&  0&  0&  0&  0\\
 449&  0&  0&  0&  0& 10& 39&100& 35&  8&  0&  0&  0&  0\\
 497&  0&  0&  0&  0&  1&  6& 35&100& 34&  5&  1&  0&  0\\
 539&  0&  0&  0&  0&  0&  0&  8& 34&100& 20&  9&  0&  0\\
 595&  0&  0&  0&  0&  0&  0&  0&  5& 20&100& 33&  4&  0\\
 621&  0&  0&  0&  0&  0&  0&  0&  1&  9& 33&100& 12&  1\\
 691&  0&  0&  0&  0&  0&  0&  0&  0&  0&  4& 12&100& 14\\
 753&  0&  0&  0&  0&  0&  0&  0&  0&  0&  0&  1& 14&100\\
\hline
$\sigma$&0.49&0.40&0.37&0.33&0.31&0.29&0.28&0.27&0.26&0.26&0.26&0.26&0.26
\end{tabular}
\end{center}
\caption{The ${\cal S}_A$ correlation matrix, $C_{AB}/\sigma_A\sigma_B$
with $\sigma_A\equiv C_{AA}^{1/2}$, for every {\it second\/} estimate of
the power spectrum from a {\it single pointing\/} of DASI.
Correlations are listed as percentages.  The final row shows the relative
error on the power spectrum in the bin.  Since the error is dominated by
sample variance, for $N$ such pointings $\sigma$ is reduced by $N^{1/2}$.
All calculations assume a scale-invariant COBE normalized spectrum.}
\label{tab:corr}
\end{table}

\end{document}